\documentclass[sigconf]{acmart}

\AtBeginDocument{%
  }

\setcopyright{acmcopyright}
\copyrightyear{2018}
\acmYear{2018}
\acmDOI{XXXXXXX.XXXXXXX}

\acmPrice{15.00}
\acmISBN{978-1-4503-XXXX-X/18/06}

\usepackage{xcolor}
\usepackage{booktabs}
\usepackage{multirow}
\usepackage{cleveref}
\usepackage[inline]{enumitem}

\begin{document}

\title{Requirements for Knowledge Graph Engineering in 2023}
\title{Consolidating the Requirements for Knowledge Engineering}
\title{Identifying and Consolidating Knowledge Engineering Requirements}

\author{Bradley P. Allen}
\affiliation{%
  \institution{University of Amsterdam}
  \city{Amsterdam}
  \country{The Netherlands}}
\email{b.p.allen@uva.nl}

\author{Filip Ilievski}
\affiliation{%
 \institution{USC Information Sciences Institute}
 \city{Marina del Rey}
 \state{CA}
 \country{USA}}
 \email{ilievski@isi.edu}

\author{Saurav Joshi}
\affiliation{%
 \institution{USC Information Sciences Institute}
 \city{Marina del Rey}
 \state{CA}
 \country{USA}}
 \email{syjoshi@isi.edu}





\renewcommand{\shortauthors}{Allen, Ilievski, and Joshi}

\begin{abstract}
Knowledge engineering is the process of creating and maintaining knowledge-producing systems. Throughout the history of computer science and AI, knowledge engineering workflows have been widely used because high-quality knowledge is assumed to be crucial for reliable intelligent agents. However, the landscape of knowledge engineering has changed, presenting four challenges: unaddressed stakeholder requirements, mismatched technologies, adoption barriers for new organizations, and misalignment with software engineering practices. In this paper, we propose to address these challenges by developing a reference architecture using a mainstream software methodology. By studying the requirements of different stakeholders and eras, we identify 23 essential quality attributes for evaluating reference architectures. We assess three candidate architectures from recent literature based on these attributes. Finally, we discuss the next steps towards a comprehensive reference architecture, including prioritizing quality attributes, integrating components with complementary strengths, and supporting missing socio-technical requirements. As this endeavor requires a collaborative effort, we invite all knowledge engineering researchers and practitioners to join us.



%
\end{abstract}




\maketitle

\section{Introduction}
\label{sec:intro}


\textit{Knowledge engineering (KE)} is the discipline of building and maintaining processes that produce knowledge. Per~\cite{ramsey1929knowledge}, knowledge can be defined as a set of beliefs that are ``(i) true, (ii) certain, (iii) obtained by a reliable process''. KE workflows have been popular throughout the evolution of computer science and AI under the intuitive assumption that the reliability of intelligent agents (e.g., chatbots) strongly depends on high-quality knowledge~\cite{ramsey1929knowledge,newell1958elements,feigenbaum1977art,feigenbaum1992personal,schreiber2000knowledge,berners2001semantic,petroni2019language,hogan2020semantic,ilievski2020kgtk,bender2021dangers,hartig2022reflections,wqdssearchteam2022wqds,alkhamissi2022review}. And yet, KE as a discipline has changed considerably since its initial flowering during the period associated with expert systems development in the nineteen-eighties.



\begin{figure*}[ht!]
    \centering
    \includegraphics[width=0.94\textwidth, trim={0 35 0 30},clip]{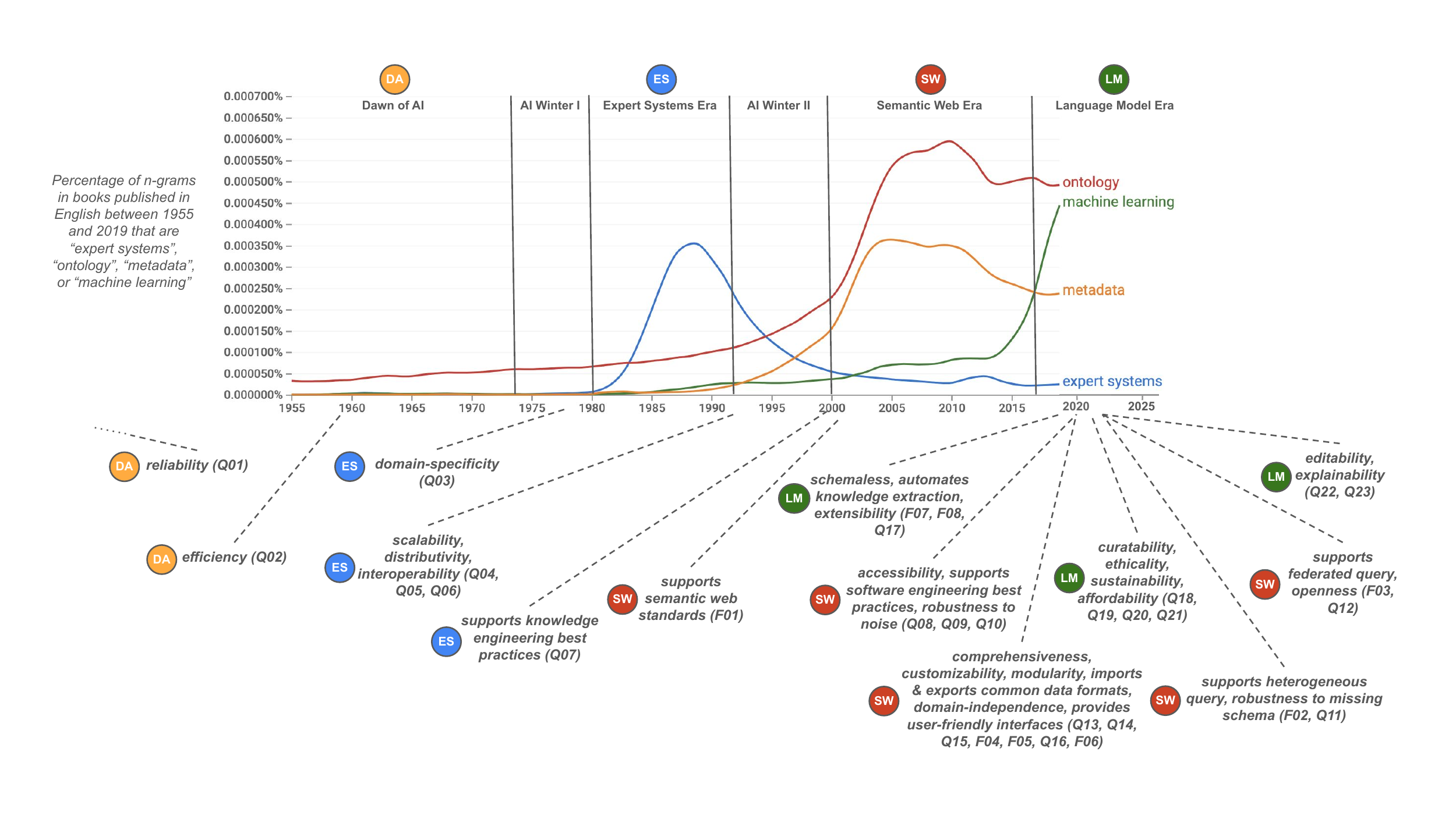}
    \caption{Evolving requirements over four eras of the history of knowledge engineering, as witnessed by research that has been responding to the shifting KE landscape~\cite{ramsey1929knowledge,newell1958elements,feigenbaum1977art,feigenbaum1992personal,schreiber2000knowledge,berners2001semantic,petroni2019language,hogan2020semantic,ilievski2020kgtk,bender2021dangers,hartig2022reflections,wqdssearchteam2022wqds,alkhamissi2022review}. We show English language book n-gram occurrence percentages over time for n-grams indicative of each era \cite{googlebooksngramviewer2022query}. We describe the figure in detail in \autoref{sec:reqs}.}
    \label{fig:kerequirementstimeline}
\end{figure*}

During the period from 1955 to today, we can identify four distinct eras in the history of knowledge engineering. We dub these the Dawn of AI, the Expert Systems Era, the Semantic Web Era, and the Language Model Era. 
\autoref{fig:kerequirementstimeline}, which we discuss in detail below, displays the four eras on a timeline, showing that the requirements for KE processes have been shifting in response to the perceived shortcomings of systems created in the preceding period.
KE has focused on reliability during the Dawn of AI, domain-specific workflows in the Expert Systems era, accessible and interoperable knowledge production in the Semantic Web era, and curatable and affordable workflows in the latest Language Model era. The dynamic requirements have been accompanied by a shift in the target stakeholders, e.g., the Semantic Web era has focused on the needs of knowledge engineers, whereas the language modeling era aims to primarily enable data scientists.
From today's perspective, the divergence of stakeholders and requirements has resulted in four key challenges. 
(1) Each era has \textbf{pain points}, i.e., requirements that have not been fully incorporated: language model engineering suffers from low interoperability~\cite{bender2021dangers}, the Semantic Web is vulnerable to the lack of high-quality schemas~\cite{hartig2022reflections}, and expert systems' pain point is scalability~\cite{feigenbaum1992personal}. 
(2) While many knowledge resources and tools have been developed, they rely on misaligned \textbf{era-specific technologies}, i.e., bespoke knowledge representation choices, domain-specific languages and systems, and siloed data-sharing practices, which has stymied progress.
(3) The lack of consensus on KE best practices \textbf{hinders adoption} for organizations wanting to apply KE to their use cases, as the costs of setting up, maintaining, and extending these stacks for organizations are unclear. Meanwhile, large internet companies such as Google and Facebook rely on custom architectures and representations that most enterprises may struggle to reproduce. (4) The KE practices adopted by different communities \textbf{do not align well with software engineering (SE)} best practices, which makes adoption even more challenging.


\textit{How does one build a standard workflow for KE that simultaneously supports the requirements of diverse stakeholders in a way that will address the above four challenges (1-4)?} We propose to  develop a \textit{reference architecture (RA)}, a key SE artifact that enables technology adoption within a community of practice by capturing design patterns that together provide a flexible framework to build and design software systems. 
RAs provide numerous advantages, such as interoperability, reduced development costs, improved communication among stakeholders, and the use of best practices, which collectively help overcome barriers such as organizational resistance and lack of expertise \cite{martinez2015aggregating}. Although there may be concerns about the learning curve associated with RAs and the need for concrete examples \cite{martinez2017benefits}, the literature agrees that RAs simplify the development process, promote standardization, and facilitate collaboration. By focusing on usability and consolidation, RAs can minimize the learning curve and provide well-documented, application-oriented processes supported by easy-to-use and well-integrated tools. These benefits facilitate the adoption of data analytics in enterprises \cite{bornstein2020emerging, nadal2017software, paakkonen2020extending}, supporting workflows similar to those in KE. 

By providing these benefits, an RA can be used to directly address the four key challenges described above. Through consolidation, RAs can \textbf{allow stakeholders to effectively communicate and collaborate (1)}, analyzing their requirements, prioritizing their choice of technologies, and detecting and addressing pain points early in the KE process. Through standardization, RAs can \textbf{harmonize different (era-specific) perspectives (2)} into a single infrastructural view. By supporting usability and interoperability, RAs can \textbf{facilitate adoption by new organizations (3)} by mapping requirements to the components or patterns in the RA that address them. And, by simplifying the development process, RAs can \textbf{facilitate alignment of KE with SE best practices (4)}.


Thus, we believe that developing an RA for KE is a critical step toward the adoption of knowledge technologies by diverse stakeholders, including the mainstream SE community. Acknowledging that the development of an RA for KE is an ambitious endeavor and one that requires community discussions, in this paper we make the first step, by identifying and consolidating requirements that have arisen across diverse stakeholders and KE eras. We start with studying RA development methodologies and analyzing the gap between existing KE practices and mainstream RA practices in SE, in~\autoref{sec:relwork}. We profile prominent KE stakeholder personas, each with their typical needs, corresponding tooling, and pain points as indicators for unsatisfied needs, in~\autoref{sec:profile}. Then, we dive deeper into the requirements of these stakeholders within the four KE eras since the dawn of AI and derive a consolidated list of requirements in \autoref{sec:reqs}. 
In \autoref{sec:evaluation}, we show how our derived requirements can be applied to evaluate the suitability of three prominent and relatively recent KE architectures as future RAs for KE. The satisfaction of the requirements can be traced to the focus of each architecture, whereas socio-technical requirements (e.g., curatability) are missed by all architectures. We suggest that future KE architectures should be evaluated against our requirements and they should attempt to address the missing requirements.  Inspired by our findings, we conclude with a discussion of the next steps toward a comprehensive RA for KE in \autoref{sec:conclusion}.
\section{Background}
\label{sec:relwork}

In this section, we
introduce prior work on RAs, describe common SE methodologies for creating RAs, and explain current practices in KE workflows. Thus, we characterize the gap between mainstream RA and KE practices and describe how this gap could be bridged. 

\noindent \textbf{Reference architectures} 
Per Angelov et al.~\cite{angelov2009classification}, an RA is a generic architecture for a class of information systems that are used as a foundation for the design of concrete architectures in this class. 
RAs provide the highest level of abstraction, they emphasize heavily architectural qualities, their stakeholders are considered but absent from the architecture, they promote adherence to common standards, and are effective for system development and communication~\cite{ataei2022state}. 
Many RAs have been proposed in the past decades, some of which have gained wide adoption in their domains~\cite{garces2021three}. 
The Andreessen Horowitz reference architecture~\cite{bornstein2020emerging} for emerging data infrastructure and platforms is a snapshot of the current industry stack and trends that subsumes most current uses of data within an enterprise. 
Typical RAs for big data usually follow a three-step lifecycle consisting of data ingestion, transformation, and serving~\cite{ataei2022state}. 
Reference architectures of this type are examples of a key element in domain-specific software architectures (DSSAs) \cite{taylor2010software}. DSSAs combine an understanding of the problem space being addressed with business goals and technology stack solutions to provide guidance in the implementation of software architectures for a given area of application.
It is noteworthy that the Andreessen Horowitz RA makes no reference at all to mainstream KE technologies, like knowledge graphs (KGs) or Semantic Web concepts or products, especially given the care it takes to address specific use cases related to machine learning.

\noindent \textbf{RA methodologies}
A method to design a software architecture by~\cite{nakagawa2011aspect} consists of five steps: establishing its scope, selecting and investigating information sources, performing an architectural analysis to identify significant requirements, synthesizing the RA, and evaluating it through surveys as well as its instantiation and use. Software architecture methodologies leverage requirements as a common denominator between stakeholder needs and technical patterns, where the features and the functions are modeled in a product-independent way~\cite{cloutier2010concept}. Two types of requirements are commonly used: functional requirements (FRs) and quality attributes (QAs)~\cite{bass2022software}. 
\textit{Functional requirements} describe what the system components are responsible for, i.e., they state what the system must do and how it must behave or react to runtime stimuli~\cite{bass2022software}. FRs are satisfied by assigning an appropriate sequence of responsibilities throughout the architectural design. 
Bass et al.~\cite{bass2022software} define a\textit{ quality attribute} as ``a measure or testable property of a system that is used to indicate how well the system satisfies the needs of its stakeholders''. QAs should be unambiguous and testable, and measure the ability of a system to satisfy the stakeholder's goals, characterized using one or more scenarios. QAs can be mapped into general categories, e.g., those that describe a runtime property of the system (e.g., availability), or those that describe a developmental property of the system (e.g., testability). Designing for a system that satisfies all QAs is a matter of making the appropriate tradeoffs~\cite{bass2022software}. In this work, we follow the RA methodology and terminology from Bass et al.~\cite{bass2022software}, and align our requirements with the categories in this book.

\noindent \textbf{KE workflows} CommonKADS~\cite{schreiber2000knowledge} is a methodology for the extraction of expert knowledge into knowledge bases based on lifecycle and corresponding models. The popular Semantic Web Stack~\cite{hendler2009tonight} prescribes a layered infrastructure of technologies and languages that would constitute a Semantic Web application. Recognizing that knowledge can be extracted at scale from unstructured data, by crowdsourcing, and by neural models, recent work has developed KE workflows that are domain- or application-specific. KGs are popular within the Library and Information Studies (LIS)~\cite{tharani2021much} community or in e-commerce applications like the Amazon Product KG~\cite{zalmout2021all}, resulting in specific architectures. This pluralism of architectures, which may or may not adhere to the Semantic Web Stack, as well as the importance of considering a broad set of KE stakeholders (e.g., practitioners beyond academia) have been recognized by Hogan~\cite{hogan2020semantic}. Hogan notes that the broad adoption of KE practices and artifacts is hindered by the lack of understanding of users, and the mismatch between existing KE approaches and mainstream SE tools and practices. Ironically, while the field of Semantic Web puts a lot of emphasis on developing artifacts like ontologies and KGs that enable common understanding between humans and machines, it has not caught up on the idea of developing shared architectures through which different concerns can be expressed, negotiated, and resolved among stakeholders for large, complex systems~\cite{bass2022software}. We address this critical gap by consolidating the requirements across stakeholders and eras, with the final aim to devise an RA for KE.

\section{Stakeholder Profiles}
\label{sec:profile}

Since an RA is an instrument for reaching a consensus between stakeholders, it is essential to understand their requirements~\cite{cloutier2010concept}. 
We briefly describe the presently addressed needs and pain points (as open challenges) of four stakeholder profiles: a knowledge explorer, a software developer, a data scientist, and a knowledge engineer. While the full accounting of an evaluation of an RA would include profiles of business or organizational stakeholders; for the purpose of this discussion, we focus on these four profiles as useful abstractions of user tasks to drive our understanding of their needs independent of a specific business or organizational application. 

\noindent \textbf{Knowledge explorers} access data to find (usually) small pieces of information from user-friendly interfaces. As such, they are not expected to be proficient with KGs, AI, or even computer science. Knowledge explorers often are domain experts, e.g., in publishing, health, or e-commerce. 
A knowledge explorer might 
visualize knowledge in intuitive ways,\footnote{\url{https://gephi.org/}} or perform semantic text search~\cite{ilievski2016lotus}. 
Pain points for knowledge explorers include interfaces with ambiguous semantics (\textbf{accessibility})~\cite{hogan2020semantic}, misaligned tooling (\textbf{comprehensiveness})~\cite{ilievski2020kgtk}, and lack of \textbf{modular} design~\cite{ilievski2020kgtk}.

\noindent \textbf{Software engineers} design, develop, maintain, test, and evaluate computer software. 
Background knowledge may benefit software engineers in several ways, e.g., through content enrichment~\cite{lyu2022achieving} or recommendation techniques~\cite{zalmout2021all}.
Software engineers face a mismatch of assumptions and common KE practices, i.e., the stack of tooling and formats used by software engineers is often different from those developed by knowledge graph engineers. Further obstacles include the lack of \textbf{robustness to a missing schema}~\cite{hartig2022reflections}, the lack of software developer-friendly documentation and data catalogs (\textbf{accessibility})~\cite{hogan2020semantic}, and the entry-level costs (\textbf{affordability})~\cite{bender2021dangers}.

\noindent \textbf{Data scientists} draw insights and build predictive models from data using various techniques, typically based on machine learning.
A common goal of data scientists is to minimize failure in prediction tasks according to metrics like accuracy and precision. Data scientists may use available knowledge to enhance their training process and ultimately boost or evaluate the performance of their models. Knowledge may be used to support the development of a model for a specific task (e.g., entity linking~\cite{ayoola2022refined}), a larger suite of tasks (e.g., analytical operations), or support benchmarking and evaluation~\cite{widjaja2022kgxboard}. Yet, the integration of knowledge structures in data science research may be hindered by the lack of \textbf{comprehensiveness} in tooling, the quality of the knowledge produced by the KE process (\textbf{robustness to noise})~\cite{ilievski2020kgtk}, and the ability to \textbf{scale} economically with the amount of knowledge produced~\cite{feigenbaum1992personal}.


\noindent \textbf{Knowledge engineers} build, maintain, and query knowledge-based systems. Consequently, they require efficient and easy-to-use ways to access and modify information systematically. Tools for knowledge engineers support querying (e.g., by graph databases like Graph DB)\footnote{\url{https://www.ontotext.com/products/graphdb}} or support contributions to knowledge artifacts (e.g., via ontology editors like Protege~\cite{noy2003protege}). Pain points include the lack of tools that support the creation of large-scale knowledge artifacts (\textbf{scalability})~\cite{feigenbaum1992personal}, missing support for manual curation of knowledge (\textbf{curatability})~\cite{bender2021dangers}, and difficulty in terms of incorporating more sources or modalities of data (\textbf{extensibility})~\cite{petroni2019language}.



\section{A historiographical approach to identifying KE requirements}
\label{sec:reqs}



We argue that a useful way to understand the evolution of stakeholders' requirements for KE is to examine representative literature from its four eras that take a specific position on what those requirements should be. 
Following mainstream SE practices and inspired by our characterization of stakeholder profiles, in this section, we derive a list of functional requirements (Table~\ref{tab:kequalattrs}) and quality attributes (Table~\ref{tab:kefuncreqs}). 
Before we present the set of FRs and QAs, we discuss two caveats. First, we note that the FRs and QAs are inherently biased by the authors' knowledge and perspective of prior work. Second, as we aim to be faithful to the perspective of multiple stakeholders, some of the requirements may be interdependent, i.e., they may partially overlap or even contradict each other. We allow for the subjectivity and the dependencies to co-exist at this point. 



\subsection{Requirements from the Dawn of AI}
We start with Frank Ramsey's 1929 definition of knowledge as a set of beliefs that are ``(i) true, (ii) certain, [and] (iii) obtained by a reliable process'' \cite{ramsey1929knowledge} as a baseline requirement that KE processes be \textbf{reliable (Q01)}. From the late nineteen-fifties, some of the earliest work in AI additionally identified the requirement that such processes also be \textbf{computationally efficient (Q02)}, in the sense that they complete execution in a reasonable amount of time and space \cite{newell1958elements}. Newell and Simon~\cite{newell1958elements} were optimistic about the potential of goal-directed search using heuristics as a general approach to problem-solving to be useful for practical applications. Still, by the beginning of the nineteen-seventies, it was clear that such systems were difficult to use in developing applications that were recognizably more than just toy tasks. 

\subsection{Requirements from the Expert Systems era}
By the mid-seventies, having been deeply involved in attempting to apply Newell and Simon's model, Feigenbaum became convinced that automating knowledge production required a \textbf{domain-specific (Q03)} focus to succeed \cite{feigenbaum1977art}. His evangelism of KE (a term he was instrumental in propagating the use of) engendered a period of intense activity in the construction of expert systems for the purposes of decision support in business enterprise settings. By the early nineteen-nineties, however, Feigenbaum and others acknowledged that the expert systems approach resulted in systems that were brittle and hard to maintain. Without abandoning his requirement that KE is domain-specific in application focus and thus heavily dependent on subject matter expertise, he argued that future knowledge-based systems also be \textbf{scalable (Q04)}, \textbf{globally distributed (Q05)}, and \textbf{interoperable with other knowledge bases (Q06)} to address these shortcomings \cite{feigenbaum1992personal}. At that point in time, there was no consensus about how such requirements could be addressed, but in retrospect, one can argue that in \cite{feigenbaum1992personal} Feigenbaum anticipated several aspects of what several years later would come to be known as the World Wide Web. As the nineties progressed, efforts were made to provide \textbf{support for KE best practices (Q07)} through the definition of structured methodologies \cite{schreiber2000knowledge}.

\begin{table*}[]
\caption{Quality attributes for knowledge engineering. We map each of the requirements to the closest category from the software engineering handbook by~\citeauthor{bass2022software} \cite{bass2022software}.}
\label{tab:kequalattrs}
\small
\begin{tabular}{l|p{2.7cm}|l|p{10.6cm}|l}
\textbf{id} & \textbf{requirement} & \textbf{ref.} & \textbf{scenario} & \textbf{category}\\ \hline
Q01 & reliability & \cite{ramsey1929knowledge} & the knowledge engineering process is reliable, i.e. the knowledge produced can be trusted to be true and justified & reliability \\
Q02 & efficiency & \cite{newell1958elements} & the knowledge produced by the KE process can be applied in a computationally tractable and efficient manner & performance \\
Q03 & domain-specificity & \cite{feigenbaum1977art} & the KE process is tailored to use cases associated with a specific domain or area of expertise & maintainability \\
Q04 & scalability & \cite{feigenbaum1992personal} & the knowledge engineering process scales economically with the amount of knowledge produced (measured in terms of, e.g. rules, triples, nodes, edges, etc.) & scalability \\
Q05 & distributivity & \cite{feigenbaum1992personal} & the knowledge produced by the KE process can be distributed and hosted across multiple sites & availability \\
Q06 & interoperability & \cite{feigenbaum1992personal} & the knowledge produced by the KE process can be easily shared across sites and applications & interoperability \\
Q07 & supports knowledge engineering best practices & \cite{schreiber2000knowledge} & the KE process 
follows methodologies for creating an ontology and elicitation of knowledge from subject matter experts towards the creation of a knowledge model & maintainability \\
Q08 & accessibility & \cite{hogan2020semantic} & the barrier to adoption by users of the KE process is low & usability \\
Q09 & supports software engineering best practices & \cite{hogan2020semantic} & the knowledge engineering process conforms to software industry norms (e.g. the use of agile methodologies, continuous integration and deployment, version control, automated testing, automated vulnerability scans, etc.) & maintainability \\
Q10 & robustness to noise & \cite{hogan2020semantic} & the KE process is robust in the face of noise and/or adversarial manipulation of source data and/or knowledge & reliability \\
Q11 & robustness to missing schema & \cite{hartig2022reflections} & the knowledge produced by the KE process can be processed and/or accessed in the face of incomplete schemas and/or knowledge organization systems & reliability \\
Q12 & openness & \cite{wqdssearchteam2022wqds} & the components of the KE process are implemented using open source software, with open standards, and the knowledge produced by the knowledge engineering process is openly accessible & maintainability \\
Q13 & comprehensiveness & \cite{ilievski2020kgtk} & all components of an end-to-end KE process (e.g. data ingest/export, data transformation, inference, knowledge publishing, etc.) are supported & usability \\
Q14 & customizability & \cite{ilievski2020kgtk} & the components of the KE process can be modified to support specific use cases & usability \\
Q15 & modularity & \cite{ilievski2020kgtk} & the components of the KE process can be selectively composed to suit a specific use case & maintainability \\
Q16 & domain-independence & \cite{ilievski2020kgtk} & the KE process is generally applicable across a wide range of domains and areas of expertise & usability \\
Q17 & extensibility & \cite{petroni2019language} & knowledge extraction  from data or natural language performed in the KE process can easily accommodate new sources and modalities of data or natural language & maintainability \\
Q18 & curatability & \cite{bender2021dangers} & the KE process supports human curation of automatically extracted and/or inferred knowledge & safety \\
Q19 & ethicality & \cite{bender2021dangers} & the KE process supports compliance with and enforcement of policies and/or guidelines for ethical use & safety \\
Q20 & sustainability & \cite{bender2021dangers} & the cost of executing the KE process is economically sustainable for the given use case & usability \\
Q21 & affordability & \cite{bender2021dangers} & the cost of access to the KE process is economically affordable for a given user community & usability \\
Q22 & editability & \cite{alkhamissi2022review} & the knowledge produced by the KE process can be feasibly edited by humans & maintainability \\
Q23 & explainability & \cite{alkhamissi2022review} & the knowledge produced by the KE process provides accountability with respect to its provenance and the details of how it was produced (e.g. through human authoring, automated extraction, and/or inference, etc.) & testability
\end{tabular}%
\end{table*}

\begin{table*}[]
\caption{Functional requirements for knowledge engineering.}
\label{tab:kefuncreqs}
\small
\begin{tabular}{l|p{3.8cm}|p{.4cm}|p{12cm}}
\textbf{id} & \textbf{requirement} & \textbf{ref.} & \textbf{scenario} \\ \hline
F01 & supports semantic web standards & \cite{berners2001semantic} & the KE process supports the use of W3C semantic web standards, including those for knowledge representation (RDF), serializations (e.g. Turtle, JSON-LD, etc.) and querying (SPARQL) \\
F02 & supports heterogeneous query & \cite{hartig2022reflections} & the knowledge produced by the KE process can be queried using multiple query languages (e.g. SQL, Cypher, SPARQL, etc.) and query execution strategies (e.g. federated query, centralized query, find-and-follow, etc.) \\
F03 & supports federated query & \cite{wqdssearchteam2022wqds} & the knowledge produced by the KE process can be queried across multiple sites\\
F04 & imports common data formats & \cite{ilievski2020kgtk} & the KE process supports import of data and/or knowledge from software industry standard data sources (e.g. relational databases, RDF data dumps, REST APIs, etc.) and serializations (e.g. CSV, JSON, Parquet, etc.) \\
F05 & exports common data formats & \cite{ilievski2020kgtk} & the knowledge engineering process supports export of produced knowledge to software industry standard data delivery mechanisms (e.g. as serialized data dumps, through publish/subscribe messaging, through REST APIs, search engine indexes, etc.) \\
F06 & provides user-friendly interfaces & \cite{ilievski2020kgtk} & the knowledge produced by the knowledge engineering process can be accessed and applied through one or more user-friendly interfaces (e.g. command line interfaces, visual editors and browsers, REST APIs, etc.) \\
F07 & schemaless & \cite{petroni2019language} & the knowledge engineering process does not need schemas to be defined to support effective knowledge extraction from data and/or natural language \\
F08 & automates knowledge extraction & \cite{petroni2019language} & knowledge extraction from data or natural language performed in the KE process is automatic and does not require human labor
\end{tabular}%
\end{table*}

\subsection{Requirements from the Semantic Web era}
With the establishment of the Web and the emergence of Web architectural principles, Berners-Lee argued for a ``Web of Data'' based on linked data principles, standard ontologies, and data-sharing protocols that not only provided an implementation of Feigenbaum's requirements, but with a single stroke established Web-centric open standards that \textbf{use W3C semantic web standards (e.g., RDF, SPARQL) (F01)} that anyone could adopt \cite{berners2001semantic}. The subsequent twenty years witnessed the development of a globally federated open linked data ``cloud'', as well as the refinement of techniques for ontology engineering, i.e., the development and publishing of shared data schemas with semantics using linked data principles. Enterprises in particular found better value propositions for using such techniques toward the improvement of access and discovery of Web content and data, in contrast to the automation of decision-making that was the primary value proposition for knowledge-based systems during the expert systems era \cite{hendler2020semantic}. However, while progress was made in building systems based on such principles, the general adoption of specific principles advocated for by the semantic web community by the broader community of software developers and web application designers was slow, leading semantic web researchers to identify additional requirements for broader adoption, such that the core tools and standards used in semantic web application be more \textbf{developer-friendly (Q08)} and more directly \textbf{aligned with software industry norms (Q09)}, and that measures be taken to make federated open data more \textbf{robust to noise in data sources (Q10)} \cite{hogan2020semantic}. Additional focus on \textbf{support for heterogeneous query methods (F02)} and \textbf{support for federated query (F03)}, and on making access to linked data \textbf{robust to data catalog incompleteness (Q11)} while maintaining the practical benefits of \textbf{open source and open standards (Q12)} led to new requirements towards those ends \cite{hartig2022reflections,wqdssearchteam2022wqds}.


In projects over the last ten years where commercially-useful enterprise KGs have been produced, such as the Google KG \cite{noy2019industry} and Amazon's Product KG \cite{dong2018challenges}, the attempt to address Q08 and Q09 has led to a reliance on custom architectures and approaches, which do not address the requirements of interoperability and federation identified by Feigenbaum and Berners-Lee \cite{feigenbaum1992personal,berners2001semantic}. The impact of these projects, together with efforts to establish architectures that improve the developer experience associated with the development of KGs, has led to the need for a finer-grained articulation of what KE should provide to developers. 

We identify the following additional requirements based on the principles explored in the development of the recent KGTK toolkit and our experience in its use \cite{ilievski2020kgtk}. As KE today impacts a wide range of disciplines and stakeholders, its infrastructures need to be \textbf{comprehensive (Q13)}, i.e., encompass the wide range of typical KE operations: data collection, processing, querying, transformation, and distribution. Standardization and adoption of KE practices rely on the presence of such comprehensive tooling, as already shown by the role of infrastructures like HuggingFace for natural language processing,\footnote{\url{https://huggingface.co}} and scikit-learn for machine learning. \footnote{\url{https://scikit-learn.org}} To make the architecture accessible and understandable for a wide range of stakeholders, KE should be familiar and relevant to the world of each stakeholder, designed to follow their best practices as closely as possible (cf. Q07 and Q09). 
KE needs to be scalable (Q04): popular KGs today, like Wikidata, 
count their statements in the billions. While modern hardware and software can support the manipulation of data of such sizes, this is a non-trivial challenge and requires solutions that can scale with the current and future size of these resources.
Given the variety of intended use cases, KE needs to be \textbf{customizable (Q14)}: its components should allow for enough flexibility to be adaptable to diverse stakeholder needs. To facilitate Q14, the infrastructure should be designed in a \textbf{modular way (Q15)} where the user can pick and choose the components that are of use to them, and ignore the rest. 
A key aspect of modular architecture is to provide stakeholders with atomic functions for common operations, such as getting labels and aliases for an item, describing an item, querying for similar items, text searching for entities or events, fact extraction, and extracting a subset for reuse. As the adoption of KE is tightly coupled with their role in broader systems, the infrastructure for KE should natively \textbf{integrate common data formats (F04)} used in AI and SE, including CSV, TSV, JSON, and SQL. KE processes should support the \textbf{production of common data formats (F05)}, analogous to those in F04. Together, F04 and F05 ensure that KE is interoperable with mainstream AI software, like HuggingFace.
KE should be \textbf{domain-independent (Q16)}, i.e., generalizable to a wide range of use cases. Ideally, a representative selection of use cases should be provided to illustrate the simplicity, effectiveness, and efficiency of including knowledge technologies in developer tasks. 
To support rapid prototyping and make KE easy to get started (cf. Q08), KE should provide \textbf{user-friendly interfaces (F06)}, both in the form of command language as well as graphical interfaces like browsers and search endpoints.


\subsection{Requirements from the Language Model era}
The success of connectionist methods arising from the proliferation of graphical processing hardware for matrix arithmetic and concurrent innovations in neural architectures \cite{wang2017origin} has led to a new set of possibilities for the production of knowledge. While at the time of this writing, it is difficult to summarize this area due to the burst of research in this direction, two perspectives on the relation between language models and knowledge bases have emerged over the last several years, both yielding new KE requirements.

One perspective is that a language model can serve directly as a knowledge base that is queryable using natural language prompts \cite{petroni2019language}. The argument is made that given this, new requirements for knowledge production are that it \textbf{can extract knowledge from text without a data schema (F07)}, that is, be driven by learning directly from unstructured natural language modeling without recourse to pre-defined knowledge organization schemes, that it \textbf{can extract knowledge from text and/or data with minimal human labor (F08)}, making it \textbf{easy to incorporate new sources of data and/or knowledge (Q17)}. 
Note that these requirements are at odds with the far more human labor-intensive processes assumed in previous eras. This stance is not without controversy; though the emergent ability of large language models to generate syntactically correct and complex natural language is on the surface impressive, approaches to ensure that the natural language produced is meaningful and correct are still under development, and commercial offerings based on large language models can often be shown to confabulate, leading to the potential for them to be used in harmful ways \cite{bender2021dangers}. This leads to new requirements that the production of knowledge using these technologies make it \textbf{easy for humans to curate extracted knowledge (Q18)} and that they \textbf{support ethical use (Q19)}. 
In addition, the cost associated with the creation of language models, and in particular the often hidden and potentially exploitative use of manual labor to prepare data to tune them, suggests the need that KE processes be \textbf{economically sustainable to run (Q20)} and \textbf{affordable to use (Q21)}. 

A more conservative perspective is that a language model can be a useful component in a KE workflow that combines techniques based on the use of language models together with more traditional symbolic approaches \cite{alkhamissi2022review}. Here the desired requirements are that language modeling be accessible to a broad community of developers and/or users (Q08), that it \textbf{supports the manual editing of extracted knowledge (Q22)} and \textbf{supports the explanation of reasoning methods (Q23)}, again reflecting a degree of concern on the harnessing of language models in ways that address the clear shortcomings of the initial generation of such technologies.

\section{Evaluating Existing Architectures}
\label{sec:evaluation}

Per Bass et al.~\cite{bass2022software}, architectures should be evaluated for the QAs they support in the context of well-defined scenarios. Accordingly, we applied Q1-23 to three architectures:  TG~\cite{tamavsauskaite2022defining}, BioCypher~\cite{lobentanzer2022democratising}, and KGTK~\cite{ilievski2020kgtk}. We have selected these three architectures because all of them attempt to provide a comprehensive approach to KE, albeit with a different focus: TG derives workflow patterns in a bottom-up manner based on a survey of existing KE methods, BioCypher proposes a generic architecture for the domain of biomedicine, whereas KGTK is a toolkit that aims to cover the comprehensive functionality needed for KE at scale. Each of these is an example of a domain-specific\footnote{Note that the sense in which the term ``domain'' is used in DSSAs is distinct from that assumed in the use of domain-specificity (Q03) and domain-independence (Q15) in the QAs. In the DSSA instance, we take ``domain'' to refer to the general practice of KE, while in the QAs listed above and the architecture evaluations below, we use this term to refer to the area of knowledge or expertise itself, independent of the practice of KE.} software architecture \cite{taylor2010software} in that they combine a reference architecture with a component library and guidance as to how to use components in the library to instantiate various elements of the reference architecture. Given an architecture, we study its design to evaluate whether it addresses a particular attribute.
Table \ref{tab:archeval} summarizes the results of this evaluation, where the QAs are grouped in the categories from~\cite{bass2022software}. 

\begin{table}[]
\caption{Evaluation of KE architectures by support for QAs. TG corresponds to Tamašauskaitė and Groth \cite{tamavsauskaite2022defining}, BC is BioCypher  \cite{lobentanzer2022democratising}, and KT is KGTK \cite{ilievski2020kgtk}. Our 23 QAs are mapped to the nine categories of~\cite{bass2022software}.}
\label{tab:archeval}
\small
\begin{tabular}{l|p{0.5cm}|p{3cm}|l|l|l}
\textbf{category} & \textbf{id} & \textbf{quality attribute} & \textbf{TG} & \textbf{BC} & \textbf{KT} \\ \hline
\multirow{3}{*}{reliability} & Q01 & reliability & \checkmark & \checkmark & \checkmark \\
 & Q10 & robustness to noise & \checkmark & \checkmark & \checkmark \\
 & Q11 & robustness to missing schema & \checkmark & - & \checkmark \\ \hline
 performance & Q02 & efficiency & \checkmark & \checkmark & \checkmark \\ \hline
 \multirow{7}{*}{maintainability} & Q03 & domain-specificity & - & \checkmark & - \\
 & Q07 & supports knowledge engineering best practices & \checkmark & - & - \\
 & Q09 & supports SE best practices & - & \checkmark & \checkmark \\
  & Q12 & openness & - & \checkmark & \checkmark \\
 & Q15 & modularity & - & \checkmark & \checkmark \\
 & Q17 & extensibility & \checkmark & \checkmark & \checkmark \\
 & Q22 & editability & - & - & - \\ \hline
 scalability & Q04 & scalability & \checkmark & \checkmark & \checkmark \\ \hline
availability & Q05 & distributivity & \checkmark & \checkmark & \checkmark \\ \hline
interoperability & Q06 & interoperability & \checkmark & \checkmark & \checkmark \\ \hline
\multirow{6}{*}{usability} & Q08 & accessibility & - & \checkmark & \checkmark \\
 & Q13 & comprehensiveness & \checkmark & - & \checkmark \\
 & Q14 & customizability & \checkmark & \checkmark & \checkmark \\
 & Q16 & domain-independence & \checkmark & - & \checkmark \\
 & Q20 & sustainability & - & \checkmark & \checkmark \\
 & Q21 & affordability & - & - & - \\ \hline
\multirow{2}{*}{safety} & Q18 & curatability & - & - & - \\
 & Q19 & ethicality & - & \checkmark & \checkmark \\ \hline
testability & Q23 & explainability & - & \checkmark & \checkmark \\
\end{tabular}%
\end{table}

\noindent \textbf{1. Reliability} maps to three quality attributes. \textbf{Reliability (Q01)} and \textbf{robustness to noise (Q10)} are satisfied by all three architectures. TG's pipeline uses inference, validation, and optimization to ensure that only valid and relevant knowledge is included in the KG. BioCypher supports reliability through end-to-end testing, violation reports, and propagation of information about evidence and provenance. KGTK supports reliability and robustness to noise through operations for validation and cleaning. TG is \textbf{robust to a missing schema (Q11)} thanks to its development of ontology to represent domain knowledge, utilizing non-relational database systems to store KGs, and leveraging machine learning to extract, process, and complete knowledge. KGTK handles missing schemas by validating that the data meets the format specification and cleaning syntactic and semantic violations of the specification. BioCypher cannot function if a schema is missing. 

\noindent \textbf{2. Performance} in the sense of \textbf{efficiency (Q02)} is emphasized in all three architectures. TG describes computationally tractable algorithms for each of its components (e.g., PageRank for the data identification component). BioCypher enables the creation of custom KGs in minutes by using its adapters for data ingestion and providing a schema configuration. KGTK focuses on large KGs and stores its intermediate results to disk for time and memory efficiency, which results in a relatively short running time. 

\noindent \textbf{3. Maintainability} is the richest category in our evaluation, as it maps to seven QAs.
Out of the three architectures, only BioCypher is \textbf{domain-specific (Q03)} as it focuses on the biomedical domain and illustrates its utility through a set of appropriate use cases. TG is not domain-specific because the same architecture is used both for manipulating a generic open KG and a domain-specific KG about user experience practices. KGTK illustrates its functionality on various domains, including scientific publications and finance, however, it is not tailored to use cases associated with any specific domain. TG \textbf{supports KE best practices (Q07)}, as it involves generic components such as ontology development, knowledge acquisition, representation, completion, validation, and maintenance. BioCypher and KGTK do not support ontology development which is a key constituent of the KE process. BioCypher and KGTK address \textbf{SE best practices (Q09)}, such as continuous integration, deployment, and testing methods. TG is more suited for initial KG development and does not support agile development, continuous integration and deployment, version control, and automated testing. \textbf{Openness (Q12)} is unclear in the case of TG, as the implementation of its components is underspecified. BioCypher is open because all its components are open-source and it is developed by a research community, while KGTK is based on open-source components with extensive documentation. \textbf{Modularity (Q15)} is supported by BioCypher through its easily replaceable adapters and KGTK through its standard data format which is used by all of its components. TG is not modular: although its components can be easily excluded, the data format may vary across its components. 
\textbf{Extensibility (Q17)} is the only QA that is explicitly addressed by all three architectures: TG includes an update step, BioCypher allows extensions through the reuse of existing adapters for new knowledge sources, while KGTK allows extensions through its join operation and import/export of common data formats. 
The aspect of \textbf{editability (Q22)} is a gap in all three architectures: each of them focuses on optimizing the automatic components and their connection, but none explains how manual editing would be carried out.

\noindent \textbf{4. Scalability} is well-supported by existing architectures. TG achieves \textbf{scalability (Q04)} by using an efficient NoSQL storage and spreading the data across a cluster of machines. BioCypher achieves scalability by separating data storage and analysis, allowing each component to be scaled individually, while using distributed computing infrastructure, such as computing clusters, to perform both tasks in close proximity. 
KGTK uses scalable SQL storage accompanied by indexing and caching mechanisms that together ensure high scalability to large KGs like Wikidata.

\noindent \textbf{5. Availability}, which corresponds to our \textbf{distributivity (Q05)}, is explicitly addressed by all three architectures. TG enables knowledge to be distributed to different sites by using clusters over a NoSQL database system, where all data is replicated for redundancy and high availability. BioCypher addresses distributivity by separating data storage from analysis and through federated machine learning. KGTK enables distributivity by providing a simple model that fits different data representations and is mapped to a database management system that supports both relational and non-relational databases.

\noindent \textbf{6. Interoperability}, i.e., the ability to share the produced knowledge across sites and applications, is supported by all three architectures. TG offers \textbf{interoperability (Q06)} via standardization of its steps to produce, complete, and store knowledge, which is finally stored in relational databases. BioCypher's ETL pipeline incorporates several resources into a single ontological framework that follows the FAIR (findable, accessible, interoperable, reusable) principles. KGTK supports interoperability by representing its knowledge with the widely used format of tab-separated values (TSV), and by storing the data internally in a relational database, which can be easily shared across applications.

\noindent \textbf{7. Usability} maps to six QAs. \textbf{Accessibility (Q08)} for novel users is addressed by BioCypher based on its FAIR principles, and by KGTK through its extensive documentation and introductory materials. As TG has no specific framework that users can install and work with, it does not address accessibility. BioCypher does not support \textbf{comprehensiveness (Q13)} as it does not contain crucial components of end-to-end KE, including knowledge acquisition and quality evaluation. Conversely, TG and KGTK aim to support all the components of the KE process. \textbf{Customizability (Q14)} is the only usability QA that is explicitly supported by all three architectures: TG and KGTK can manipulate both large generic KGs and domain-specific KGs showing they can be adopted for multi-purpose data science operations over KGs by modifying several components in its pipeline, while BioCypher supports customization through manipulating its underlying ontology with extra information. 
BioCypher does not support \textbf{domain-independence (Q16)} as it does not focus on integrating generic KGs but rather focuses on domain-specific knowledge. TG and KGTK manipulate generic KGs and domain-specific KGs achieving domain independence.
The related attribute of economic \textbf{sustainability (Q20)} is supported by BioCypher through its automatic ontology construction and curation techniques, and similar in KGTK through its validation, cleaning, and further transformation operations that anticipate no cost for manual curation.
Ontologies in TG are constructed manually which can be expensive and labor-intensive. 
However, \textbf{affordability (Q21)} is not addressed by any of the three architectures: as TG has no specific framework, its start-up cost cannot be computed; BioCypher and KGTK are open-source projects and are freely available, which may incur costs for their infrastructural setup, training, or support.

\noindent \textbf{8. Safety} is translated into two QAs. While all three architectures support distilling valid and relevant knowledge automatically and aim to minimize the need for human curation, \textbf{curatability (Q18)} by humans is not addressed by any of them.
\textbf{Ethicality (Q19)} is unclear in the case of TG since its external algorithms 
may or may not ensure legal compliance, collaborative and transparent development, community oversight, and quality control for ethical use. As an open-source project, BioCypher ensures these properties, and it does not allow the creation of entities without associated source, license, and version parameters. Similarly, KGTK's ethicality is supported by following open-source practices.

\noindent \textbf{9. Testability} maps to \textbf{explainability (Q23)}. Although each component of TG uses an efficient machine learning algorithm that incorporates feedback to evaluate the KG, TG does not include different stakeholders or clear documentation steps, and some of its algorithms may be non-interpretable. BioCypher explicitly records the evidence (which experiment and publication the knowledge is derived from) and provenance (who provided which aspects of the primary data) during the KG creation, making it accountable and explainable. KGTK's knowledge-creation process stores the intermediate outputs on disk, enabling reproducibility and inspection.

\noindent \textbf{Lessons learned} We highlight four key lessons from this evaluation. (1) Our \textbf{23 derived quality attributes can be effectively applied for a formal assessment and comparison of KE architectures}. We propose that future KE developers should keep Q1-23 in mind when devising novel architectures. (2) None of the three architectures ticks all the boxes, yet, their \textbf{requirement satisfaction aligns well with the focus} of the architectures, e.g., KGTK puts a strong focus on usability requirements, whereas BioCypher excels in terms of maintainability. (3) \textbf{Failing to satisfy a QA comes with a measurable price} - e.g., the fact that BioCypher strictly requires a schema makes it non-applicable to use cases without a high-quality schema. Meanwhile, since all architectures fare well on scalability, we expect them to be applicable to large KGs like Wikidata. (4) \textbf{Certain QAs: editability, curatability, and affordability are missed by all architectures}, confirming Hogan's~\cite{hogan2020semantic} argument for better socio-technical integration in KE, which should be addressed in future KE architectures.

\section{From Requirements To A Reference Architecture for KE?}
\label{sec:conclusion}
In our work so far, we identified the needs of various stakeholders and KE eras and consolidated them into quality attributes and functional requirements.
We demonstrated the utility of these QAs to test the suitability of current (and future) RAs for KE.
The evaluation of three candidate architectures revealed that none of the architectures satisfies the full set of QAs, though the KGTK architecture came the closest. We observed that the three architectures took different approaches to support a certain attribute (e.g., scalability), whereas attributes that relate to human involvement through curation or affordability were generally absent from all three architectures. We proposed that future KE architectures should be evaluated against these 23 QAs and 8 FRs, and they should provide better support for socio-technical requirements.

Encouraged by these insights, we plan to continue the development of an RA for KE by following standard software methodologies. We will survey relevant stakeholders for their prioritization of attributes to understand their relevance across applications. Next, we will use this prioritization to guide the creation of an RA that will combine the complementary strengths of different existing architectures. We will apply our QAs to assess additional KE architectures, such as the Wikidata ecosystem. We will also extend our evaluation to representative data engineering architectures, which will enable us to qualify the material differences between KE and data engineering. We anticipate that such a human-centric iterative methodology will yield a comprehensive RA for KE that supports the requirements of different stakeholders and KE eras, thus naturally facilitating adoption. As this endeavor requires a collaborative effort, we invite the entire KE community to join us.

\section{Acknowledgments}

We thank Nenad Medvidovic for his invaluable review and suggestions. We thank Jay Pujara and Juan Sequeda for the inspiring discussions. Filip Ilievski and Saurav Joshi are funded by the DARPA Knowledge Management at Scale and Speed (KMASS) program.

\bibliographystyle{ACM-Reference-Format}
\bibliography{ke}

\appendix 
\section{Appendices}

\subsection{Tamašauskaitė and Groth's knowledge graph development process}

\textbf{Reliability (Q01)} - If the desired knowledge graph is generic, then it covers multiple domains, and is publicly available. Otherwise, if it is domain-specific, then it is commonly used in organizations for their operations. In either case, the data collected from various sources may not be of good quality. After performing the Extract knowledge step from the architecture, the next step is Process Knowledge where one of the sub-steps is Complete Knowledge which focuses on enriching the knowledge in the knowledge graph as well as improving the overall quality. This includes performing reasoning and inference, validating the triples, and optimizing the knowledge graph. Basically, it makes sure that only valid and relevant knowledge is included in the knowledge graph. Additionally, removing nodes that are not relevant to the domain and eliminating conflicts and gaps in the knowledge graph. In summary, the knowledge produced by the knowledge engineering process can be trusted to be true and justified.

\textbf{Efficiency (Q02)} - The pipeline aims at guiding developers to build knowledge graphs which means that developers have the ability to make decisions, and choose the desired and most efficient algorithms in each step of the KG development process. For example, in
Identify Step - For data acquisition, efficient algorithms like Focused crawling, Page Rank Algorithm, and A* Algorithm can be used. For database harvesting, data mining techniques like K-means clustering, Expectation-Maximization Algorithm, and kNN can be utilized.
Extract Knowledge - In order to extract entities, techniques such as named-entity recognition (NER), sequence labeling, and word embeddings can be utilized. Similarly, in order to extract relations and attributes, methods such as neural information extraction, open information extraction, and word embeddings can be utilized.
Process Knowledge - ML models (Bert, Ensemble learning, CRFs), feature vector similarity methods, and sorted neighborhoods are some efficient methods that can be used for entity resolution in order to remove duplicates and eliminate ambiguity.
In summary, knowledge produced by the pipeline can be applied in a computationally tractable and efficient manner.

\textbf{Domain specificity (Q03)} - The Tamašauskaitė and Groth knowledge graph development process is evaluated with 2 different types of KG - generic open KG (DBpedia) and domain-specific KG (User Experience Practices Knowledge Graph) to see to what extent the process is suitable and relevant to real-life scenarios. As a result, both the KG’s process is similar to the proposed architecture. Hence, the knowledge engineering process is not tailored to use cases associated with a specific domain or area of expertise but rather supports both generic open and domain-specific KG. 

\textbf{Scalability (Q04)} - The pipeline consists of processing steps such as - extracting knowledge, processing knowledge, and constructing a knowledge graph. The data collected from the data acquisition component can be stored in a NoSQL database system as they are capable of storing and processing big data. Scalability is achieved by spreading the storage of data and the work to process the data over a large cluster of machines. Hence, the knowledge engineering process scales economically with the amount of knowledge produced (measured in terms of, e.g. rules, triples, nodes, edges, etc.)

\textbf{Distributivity (Q05)} - The pipeline consists of processing steps such as - extracting knowledge, processing knowledge, and constructing a knowledge graph. The data collected from the data acquisition component can be stored in a NoSQL database system as it is a distributed system where several machines work together in clusters. Each piece of data is replicated over those machines to deliver redundancy and high availability. Hence, the knowledge produced by the knowledge engineering process can be distributed and hosted across multiple sites.

\textbf{Interoperability (Q06)} - The key to interoperability is standardization. Steps such as process knowledge, complete knowledge, and storing KG in a relational database system ensure efforts to achieve standardization. The knowledge produced by the knowledge engineering process can be easily shared across sites and applications if the data is stored in relational databases. Whereas for non-relational databases, the process of migration can be more complex as they have different data models and query languages. Hence, the knowledge produced by the knowledge engineering process can be easily shared across sites and applications.

\textbf{Supports knowledge engineering best practices (Q07)} - The methodology adheres to standard knowledge engineering methodology as it involves the usage of domain experts to construct ontology, is a structured approach in terms of knowledge acquisition, representation, validation, and implementation, and can be tailored to specific use cases, ensures knowledge quality, ensures maintenance of the KG, and involves usage of appropriate tools and techniques to construct KG. While mapping/constructing the ontology, the pipeline makes sure that the good practices of ontology development are followed. Hence, the knowledge engineering process adheres to the methodologies for creating an ontology and elicitation of knowledge from subject matter experts towards the creation of a knowledge model.

\textbf{Accessibility (Q08)} - Although the steps in the proposed architecture are provided in a very well-structured, clear format, it has no framework that users can install and work with, due to which it does not address accessibility. Hence, the barrier to adoption by users of the knowledge engineering process is high.

\textbf{Supports software engineering best practices (Q09)} - The proposed process is more suited for initial knowledge graph development, where it is necessary to determine the data and the structure of the knowledge graph rather than the continuous development of knowledge graphs. Additionally, the process does not support metadata management, temporal aspects, versioning, and incremental updates. Hence, the knowledge engineering process does not conform to software industry norms (e.g. the use of agile methodologies, continuous integration, and deployment, version control, automated testing, automated vulnerability scans, etc.)

\textbf{Robustness to noise (Q10)} - In the Identify Data step, the authors highlight the importance of identifying the data sources as it influences the overall KG development process. Additionally, they also extract knowledge, process knowledge, and finally construct the KG by integrating and completing knowledge. This includes performing reasoning and inference, validating the triples, and optimizing the knowledge graph. Hence, the knowledge production process is robust in the face of noise and/or adversarial manipulation of source data and/or knowledge.

\textbf{Robustness to missing schema (Q11)} - As the architecture involves the development of ontology to represent domain knowledge, utilizing non-relational database systems to store KG, and leveraging machine learning to extract and process knowledge i.e to complete knowledge, it is more robust to the missing schema. Hence, the knowledge produced by the knowledge engineering process can be processed and/or accessed in the face of incomplete schemas and/or knowledge organization systems.

\textbf{Openness (Q12)} - The implementation of the algorithms utilized in the various components of the architecture is not specified. It is possible that these algorithms may or may not be open-source. Hence, the components of the knowledge engineering process are mostly not implemented using open-source software, with open standards, making the knowledge produced by the knowledge engineering process not openly accessible.

\textbf{Comprehensiveness (Q13)} - The main sections and the sub-sections of the proposed architecture include Identify Data, Construct KG Ontology, Extract Knowledge (Extract entities, extract relations, extract attributes), Process Knowledge (Integrate Knowledge, Construct or Map ontology, Complete Knowledge), Construct Knowledge Graph (Store KG, Display KG, Enable use), and Maintain KG (Evaluate KG, Update KG). Hence, all components of an end-to-end knowledge engineering process (e.g. data ingest/export, data transformation, inference, knowledge publishing, etc.) are supported.

\textbf{Customizability (Q14)} - While evaluating the proposed architecture against DBpedia and User Experience Practices Knowledge Graph, several components were modified to support specific use cases. Data were identified from various Wikimedia project sources for DBpedia, and using Google Forms sources for UEPKG. DBpedia extracts knowledge using DBpedia Information Extraction Framework (DIEF), whereas UEPKG utilizes ETL strategy to extract knowledge. The constructed KG in DBpedia is evaluated using community reviews, contributions, and feedback whereas UEPKG is evaluated based on user input, traffic analytics, and search analytics. Hence, the components of the knowledge engineering process can be modified to support specific use cases.

\textbf{Modularity (Q15)} - Although the architecture allows composing modules i.e breaking the pipeline down into smaller, independent components or modules, it does not support a simple representation format that all modules in the architecture operate on, and does not include a comprehensive set of features - import and export modules for a wide variety of KG formats, compute embeddings module, and compute graph statistics. Hence, the components of the knowledge engineering process cannot be selectively composed to suit a specific use case.

\textbf{Domain-independence (Q16)} - The proposed architecture is evaluated with 2 different types of KG - generic open KG (DBpedia) and domain-specific KG (User Experience Practices Knowledge Graph) to see to what extent the process is suitable and relevant to real-life scenarios. As a result, both the KG’s process is similar to the proposed architecture. Hence, the knowledge engineering process is generally applicable across a wide range of domains and areas of expertise.

\textbf{Extensibility (Q17)} - Update the Knowledge Graph step in the Maintain the Knowledge Graph component elaborates upon updating the KG when there is a new data source relevant to the knowledge domain. But once, the new data is identified, the process is repeated from step 1 of the architecture. In summary, knowledge extraction from data or natural language performed in the knowledge engineering process can easily accommodate new sources and modalities of data or natural language.

\textbf{Curatability (Q18)} - After performing the Extract knowledge step from the Tamašauskaitė and Groth knowledge graph development process, the next step is Process Knowledge where one of the sub-steps is Complete Knowledge which focuses on enriching the knowledge in the knowledge graph as well as improving the overall quality. This includes performing reasoning and inference, validating the triples, and optimizing the knowledge graph. Basically, it makes sure that only valid and relevant knowledge is included in the knowledge graph. Additionally, removing nodes that are not relevant to the domain and eliminating conflicts and gaps in the knowledge graph. As manual curation is not explicitly mentioned, the knowledge engineering process does not support the human curation of automatically extracted and/or inferred knowledge.

\textbf{Ethicality (Q19)} - The external algorithms utilized in each component of the architecture may or may not ensure legal compliance, collaborative and transparent development, community oversight, and quality control for ethical use. Hence, the knowledge engineering process does not support compliance with and enforcement of policies and/or guidelines for ethical use.

\textbf{Sustainability (Q20)} - The 2nd step of the process is Construct the KG Ontology that provides a top-level structure of the knowledge graph where ontologies are constructed if the domain is narrow and this can be expensive and labor intensive. The next steps in the process are Extract Knowledge, and Process Knowledge which utilize machine learning models i.e large language models to extract and process knowledge. Large language models require large amounts of data to perform well and again this would involve labor and ultimately be expensive. Hence, the cost of executing the knowledge engineering process is not economically sustainable for the given use case.

\textbf{Affordability (Q21)} - As the proposed architecture does not have a dedicated framework, the start-up cost cannot be computed and ultimately cannot be evaluated against various affordability factors such as infrastructural setup, training, or support. Hence the cost of access to the knowledge engineering process is not economically affordable for a given user community.

\textbf{Editability (Q22)} - The Process Knowledge step ensures that the knowledge extracted is of high quality. Although automated techniques are used to perform knowledge integration, eliminating redundancy, contradiction, and ambiguity, it is possible that these models might be inaccurate based on multiple factors and hence not perform well. As the manual editing feature is not explicitly mentioned, the knowledge produced by the knowledge engineering process cannot be feasibly edited by humans.

\textbf{Explainability (Q23)} - Although each component in the process uses machine learning algorithms that are simple and efficient, incorporating feedback to evaluate the KG, it does not incorporate different stakeholders, or clear documentation steps, and also some of the algorithms might be black-box and non-interpretable. Hence, it does not provide accountability for provenance and the details of how it was produced, it is not explainable.

\subsection{BioCypher}

\textbf{Reliability (Q01)} - BioCypher enables the creation of task-specific KG using a modular approach where the KG depends on consistent and comprehensive annotations of major actors in the biomedical community. The automated end-to-end testing of millions of entities and relationships per KG increases trust in the consistency of the data. Additionally, the BioCypher migration is tested end-to-end, including deduplication of entities and relationships as well as verbose information on violations of the desired structure (e.g., due to inconsistencies in the input data). Lastly, during the creation of KG, evidence (which experiment and publication the knowledge is derived from) and provenance (who provided which aspects of the primary data) are always propagated. Hence, the knowledge engineering process is reliable, i.e. the knowledge produced can be trusted to be true and justified.

\textbf{Efficiency (Q02)} - The translation framework of BioCypher is very fast and makes it easy to create custom knowledge graphs by using adapters for data ingestion and a schema configuration for graph structure and ontology mappings. With an existing configuration, it takes only a few minutes to build a knowledge graph that is specific to a task, and even starting from scratch can be done in just a few days of work. This allows for rapid prototyping and automated machine learning (ML) pipelines that iterate the KG structure to optimize predictive performance. In order to achieve high performance, property graph database technology is implemented that provides an intuitive query interface. All in all, the knowledge produced by the knowledge engineering process can be applied in a computationally tractable and efficient manner.

\textbf{Domain specificity (Q03)} - BioCypher is a FAIR (findable, accessible, interoperable, reusable) framework that transparently builds biomedical KGs while preserving the provenances of the source data. The framework demonstrates usefulness through the use cases that focus mainly on the maintenance of task-specific knowledge stores, interoperability between biomedical domains, to on-demand building of task-specific knowledge graphs for federated learning. The framework does not focus on integrating generic open knowledge but rather focuses on domain-specific knowledge. Hence, the knowledge engineering process is tailored to use cases associated with a specific domain or area of expertise.

\textbf{Scalability (Q04)} - As the amount of biomedical data grows larger, integrated analysis pipelines become more extensive and, consequently, more expensive. To ensure the success of various systems biomedicine projects, it is important to have a flexible approach to managing and analyzing large knowledge sets. BioCypher achieves scalability by separating data storage and analysis, allowing each component to be scaled individually, while using distributed computing infrastructure, such as computing clusters, to perform both tasks in close proximity. It facilitates the maintenance of Sherlock software that along with a configuration enables the project database to be upscaled to an arbitrary number of nodes on an in-house or commercial cluster just as the project requires. Hence, the knowledge engineering process scales economically with the amount of knowledge produced (measured in terms of, e.g. rules, triples, nodes, edges, etc.)

\textbf{Distributivity (Q05)} - BioCypher achieves distributivity by separating data storage and analysis, allowing each component to be scaled individually, while using distributed computing infrastructure, such as computing clusters, to perform both tasks in close proximity. Additionally, BioCypher facilitates federated machine learning by providing an unambiguous blueprint for the process of mapping input data to ontology. After organizers determine the schema for a particular machine learning project, the BioCypher schema setup can be distributed, guaranteeing consistent database organization across all training instances. Hence, the knowledge produced by the knowledge engineering process can be distributed and hosted across multiple sites.

\textbf{Interoperability (Q06)} - BioCypher is an ETL pipeline with a focus on interoperability in biomedicine. It is a FAIR (findable, accessible, interoperable, reusable) framework to transparently build biomedical knowledge graphs while mapping knowledge onto biomedical ontologies serving the purpose of achieving harmonization. BioCypher framework incorporates multiple resources such as OmniPath, CKG, CROssBAR v2, Bioteque, and Dependency Map KG. The mapping of each of these knowledge collections onto the same ontological framework ultimately ensures interoperability across various biomedical domains. Hence, the knowledge produced by the knowledge engineering process can be easily shared across sites and applications.

\textbf{Supports knowledge engineering best practices (Q07)} - BioCypher facilitates the harmonization of datasets using ontology mapping and also provides ways of updating the ontology but does not focus on ontology development. Secondly, the process does not mainly focus on knowledge acquisition which is a key constituent in the knowledge engineering process. Hence, the knowledge engineering process does not adhere to the methodologies for creating an ontology and elicitation of knowledge from subject matter experts towards the creation of a knowledge model.

\textbf{Accessibility (Q08)} - BioCypher is a framework that adheres to the FAIR principles, meaning it is designed to be findable, accessible, interoperable, and reusable. It is utilized to construct biomedical knowledge graphs in a transparent manner. By mapping the knowledge onto biomedical ontologies, BioCypher ensures harmonization, promotes human and machine readability, and facilitates access to non-specialist researchers. It increases accessibility to the community by creating user-friendly interfaces using open standards. They explicitly mention there is no framework that provides easy access to state-of-the-art KGs to the “average” biomedical researcher, a gap that BioCypher aims to fill. In summary, the barrier to adoption by users of the knowledge engineering process is low.

\textbf{Supports software engineering best practices (Q09)} - The sustainability of research software is closely tied to the level of adoption and contributions from the community. BioCypher is open-source software that employs modern continuous integration and deployment methods, and it has a diverse community of researchers and developers involved from the outset. This approach enables the creation of resilient workflows that are thoroughly tested end-to-end, ensuring the integrity of the scientific data. Hence, the knowledge engineering process conforms to software industry norms (e.g. the use of agile methodologies, continuous integration and deployment, version control, automated testing, automated vulnerability scans, etc.)

\textbf{Robustness to noise (Q10)} - BioLink and BioRegistry are 2 important pieces of BioCypher. BioLink is a framework that is used to represent biomedical concepts and their relationships. BioRegistry is another resource that provides a registry of consistent vocabularies for these concepts and also offers validation of identifiers. Additionally, performing automated testing that covers all aspects of a knowledge graph, including millions of entities and relationships, boosts confidence in the consistency of the data. Hence, the knowledge production process is robust in the face of noise and/or adversarial manipulation of source data and/or knowledge.

\textbf{Robustness to missing schema (Q11)} - BioCypher uses a BioLink model as a comprehensive and generic biomedical ontology. In cases where it may be necessary, this ontology can be substituted with more targeted and purpose-driven ontologies or enhanced to better align with the specific needs of a particular task. It uses adapters to create task-specific KG. If for a resource, an adapter doesn’t exist, it uses a schema configuration file to mediate between the structure of the input data and the resulting BioCypher KG structure. As the presence of a schema configuration and the absence of mentioning of non-relational database systems makes it difficult to handle missing schema. Hence, the knowledge produced by the knowledge engineering process cannot be processed and/or accessed in the face of incomplete schemas and/or knowledge organization systems.

\textbf{Openness (Q12)} - BioCypher is open-source software that employs modern techniques of continuous integration and deployment. From the outset, it has attracted a diverse group of researchers and developers who collaborate on the development of the project. The integration approach includes a systematic and thorough biomedical ontology, known as the BioLink model, as well as a comprehensive registry and resolver for biomedical identifier resources, referred to as the BioRegistry. Both projects, like BioCypher, are open-source and community-driven. Hence, the components of the knowledge engineering process are implemented using open-source software, with open standards, and the knowledge produced by the knowledge engineering process is openly accessible.

\textbf{Comprehensiveness (Q13)} - BioCypher is an extract-transform-load pipeline that emphasizes interoperability within the field of biomedicine. Although it incorporates processes such as knowledge representation, knowledge implementation, and knowledge maintenance, it does not explicitly include knowledge acquisition and knowledge evaluation. All in all, all components of an end-to-end knowledge engineering process (e.g. data ingest/export, data transformation, inference, knowledge publishing, etc.) are not supported.

\textbf{Customizability (Q14)} - The authors utilize the ontology manipulation capabilities offered by BioCypher to extend certain branches of the broad but fundamental BioLink ontology, where it is advantageous to have more detailed information about the data that is integrated into the knowledge graph. For example in the Tumour Board use case, the schema of BioLink has a single, general "sequence variant" class, which does not provide much detail. To address this, a detailed subtree from the Sequence Ontology (SO) to the BioLink ontology at this node was added, resulting in a hybrid ontology that combines the generality of BioLink with the precision of a specialized sequence variant ontology. Additionally, BioCypher enables the creation of a subset of the entire knowledge collection in a fast and straightforward manner, ensuring that sensitive, irrelevant, or unlicensed data is not included. In summary, the components of the knowledge engineering process can be modified to support specific use cases.

\textbf{Modularity (Q15)} - BioCypher uses a modular approach for maintaining multiple task-specific knowledge graphs (KGs) from overlapping primary resources. It involves recombining individual data "adapters" for primary resources in a reusable manner, which simplifies maintenance by allowing centralized management of each adapter, rather than requiring primary resource maintenance within each individual KG. Additionally, the main goal is to achieve KG standardization and does include a comprehensive set of features - import and export modules for a wide variety of KG formats, and compute embedding modules. Hence, the components of the knowledge engineering process can be selectively composed to suit a specific use case.

\textbf{Domain-independence (Q16)} - BioCypher is a FAIR (findable, accessible, interoperable, reusable) framework that transparently builds biomedical KGs while preserving the provenances of the source data. The framework demonstrates usefulness through the use cases that focus mainly on the maintenance of task-specific knowledge stores, interoperability between biomedical domains, to on-demand building of task-specific knowledge graphs for federated learning. The framework doesn’t focus on integrating generic open knowledge but rather focuses on domain-specific knowledge. Hence, the knowledge engineering process is not generally applicable across a wide range of domains and areas of expertise.

\textbf{Extensibility (Q17)} - The modular structure of BioCypher offers a significant benefit of reusing already existing adapters for primary or secondary knowledge sources. In the event that an adapter is not present for a specific resource, a new adapter can be created by following the pattern of an existing adapter within the framework. Additionally, the authors utilize the ontology manipulation capabilities offered by BioCypher to extend certain branches of the broad but fundamental BioLink ontology, where it is advantageous to have more detailed information about the data that is integrated into the knowledge graph. Hence, knowledge extraction from data or natural language performed in the knowledge engineering process can easily accommodate new sources and modalities of data or natural language.

\textbf{Curatability (Q18)} - From the Tumour Board use case, it is quite visible that the current manual workflow for identifying actionable genetic variants involves complex database queries to various established cancer genetics databases, which is a complex and time-consuming process. After each query, the results require curation by geneticists to ensure that there are no duplicates from different databases and to evaluate their biological relevance, level of evidence, and potential for actionability. BioCypher transforms each individual primary resource into a task-specific, integrated knowledge graph (KG). During the build process, it maps the contents of each primary resource to ontological classes, which greatly reduces the need for manual curation and harmonization of the database results. As mentioned it reduces the need for manual creation and does not explicitly mention the usage of manual curation, the knowledge engineering process does not support human curation of automatically extracted and/or inferred knowledge.

\textbf{Ethicality (Q19)} - The “strict mode” of BioCypher does not allow the creation of entities without associated source, license, and version parameters. As a result, BioCypher can prevent the redistribution of data whose original license does not permit it, and ensure that proper credit is given to the creators of the data. In short, a requirement in BioCypher for these parameters ensures that data is used in an ethical and legal manner. BioCypher is free software under an MIT license, openly developed and available. As it is an open-source project, it ensures legal compliance, collaborative and transparent development, community oversight, and quality control. These measures provide fairness, transparency, privacy, and accountability. Hence, the knowledge engineering process supports compliance with and enforcement of policies and/or guidelines for ethical use.

\textbf{Sustainability (Q20)} - BioCypher facilitates the harmonization of datasets using ontology mapping and also provides ways of updating the ontology but does not focus on ontology development using a manual process or a domain expert. Additionally, during the build process, it maps the contents of each primary resource to ontological classes, which greatly reduces the need for manual curation and harmonization of the database results. Also, it does not involve expensive labor to create huge datasets for large language models as it uses methods such as i.e BioLink, BioRegistry, etc throughout the process. Lastly, as BioCypher utilizes Sherlock, a project database can be easily scaled up to any number of nodes as needed, which not only saves computing time but also helps to reduce costs. Hence, the cost of executing the knowledge engineering process is economically sustainable for the given use case.

\textbf{Affordability (Q21)} - BioCypher is an open-source project and depending upon the use case, the start-up cost may vary, and it may incur costs for infrastructural setup, training, or support. Hence the cost of access to the knowledge engineering process is not economically affordable for a given user community.

\textbf{Editability (Q22)} - BioCypher ensures automated end-to-end testing of millions of entities and relationships per KG increasing trust in the consistency of the data. Additionally, during the build process, it maps the contents of each primary resource to ontological classes, which greatly reduces the need for manual curation and harmonization of the database results. As editing of knowledge produced is not explicitly mentioned in the paper hence the knowledge produced by the knowledge engineering process cannot be feasibly edited by humans.

\textbf{Explainability (Q23)} - BioCypher is an extract-transform-load pipeline with a focus on interoperability in biomedicine. It facilitates the harmonization of datasets using ontology mapping and ensures automated end-to-end testing of millions of entities and relationships per KG, i.e., no manual interference is involved throughout the pipeline. Secondly, during the creation of KG, evidence (which experiment and publication the knowledge is derived from) and provenance (who provided which aspects of the primary data) are always propagated. Lastly, it is an open-source project and community-driven. All in all, the knowledge produced by the knowledge engineering process provides accountability with respect to provenance and the details of how it was produced (e.g. through human authoring, automated extraction, and/or inference, etc.)

\subsection{KGTK}

\textbf{Reliability (Q01)} - The validation and clean operations under the KGTK graph curation and transformation module ensure that the knowledge produced is reliable. For example, the validate operation ensures that the data meets the KGTK file format specification, detecting errors such as nodes with empty values, values of unexpected length (either too long or too short), potential errors in strings (quotation errors, incorrect use of language tags, etc.), incorrect values in dates, etc. Similarly, the clean operation fixes a substantial number of errors detected by validate operation, by correcting some common mistakes in data encoding (such as not escaping ‘pipe’ characters), replacing invalid dates, normalizing values for quantities, languages, and coordinates using the KGTK convention for literals. Finally, it removes rows that still do not meet the KGTK specification (e.g., rows with empty values for required columns or rows with an invalid number of columns). Hence, the knowledge engineering process is reliable, i.e. the knowledge produced can be trusted to be true and justified.

\textbf{Efficiency (Q02)} - In a data-centric KG pipeline, a lot of tools are required to complete the process efficiently. If performed separately, there can be potential issues such as - some tools might not work with large KG, and interoperating between tools require data transformation scripts. KGTK is a comprehensive library of tools and methods to enable easy composition of KG operations (validation, filtering, merging, centrality, text embeddings, etc.) to build knowledge-based AI applications and as it stores intermediate results to disk for time and memory efficiency, it results in a much shorter running reported time compared to other tools. It has a specific file format, varied transformation operations, etc which helps to complete the data-science task very efficiently. To test the efficiency of KGTK, a test was performed that filters out all the Qnodes which have P31 property in Wikidata. This process took 20 hours in  Apache Jena\footnote{https://jena.apache.org/} and RDFlib\footnote{https://rdflib.readthedocs.io/en/stable/}, 4 hours and 15 min in Graphy\footnote{https://graphy.link/}, and just 1 hour and 30 min using KGTK. Hence, the knowledge produced by the knowledge engineering process can be applied in a computationally tractable and efficient manner.

\textbf{Domain specificity (Q03)} - KGTK is a data science-centric approach designed to represent, create, transform, enhance, and analyze KGs. The authors have illustrated the importance of integrating, and manipulating large open generic KGs such as Wikidata, DBpedia, and ConceptNet. Additionally, they used KGTK with the CORD-19 dataset provided by the Allen Institute for AI. They enhanced the dataset with KGs such as DBpedia, and Wikidata to incorporate gene, chemical, disease, and taxonomic information, and computing network analytics on the resulting graphs. KGTK illustrates functionalities in various domains, including scientific publications and finance, however, it is not tailored to use cases associated with any specific domain.

\textbf{Scalability (Q04)} - KGTK has illustrated usage by working with large KGs such as Wikidata, DBpedia, and ConceptNet. The KGTK format does not differentiate between attributes or qualifiers of nodes and edges and full-fledged edges. Instead, tools working with KGTK graphs can interpret edges in different ways as needed. In the KGTK file format, any element can be considered a node, and every node can have any type of edge connecting it to any other node. This allows KGTK files to be mapped to DBMS, for example, relational or non-relational databases, and can be utilized to achieve scalability. KGTK uses scalable SQL storage accompanied by indexing and caching mechanisms that together ensure high scalability to large KGs like Wikidata. Hence, the knowledge engineering process scales economically with the amount of knowledge produced (measured in terms of, e.g. rules, triples, nodes, edges, etc.)

\textbf{Distributivity (Q05)} - KGTK has illustrated usage by working with large KGs such as Wikidata, DBpedia and ConceptNet. The KGTK format does not differentiate between attributes or qualifiers of nodes and edges, and full-fledged edges. Instead, tools working with KGTK graphs can interpret edges in different ways as needed. In the KGTK file format, any element can be considered a node, and every node can have any type of edge connecting it to any other node. This allows KGTK files to be mapped to DBMS, for example, relational or non-relational databases, and can be utilized to achieve distributivity. Hence, the knowledge produced by the knowledge engineering process can be distributed and hosted across multiple sites.

\textbf{Interoperability (Q06)} - Interoperating between tools in a data science-centric KG construction pipeline may require developing data transformation scripts as some of them may not support the same input/output representation. KGTK uses a standard KGTK file format which is tab-separated values (TSV) to represent edge lists, making it easy to process with many off-the-shelf tools. Additionally, The knowledge produced by the knowledge engineering process can be easily shared across sites and applications if the data is stored in relational databases. Whereas for non-relational databases, the process of migration can be more complex as they have different data models and query languages. Hence, the knowledge produced by the knowledge engineering process can be easily shared across sites and applications.

\textbf{Supports knowledge engineering best practices (Q07)} - KGTK is a framework for manipulating, validating, and analyzing large-scale KGs. The pipeline mechanism involves multiple modules such as data ingestion, data integration, data manipulation, data validation, data cleaning, querying, analytics, and export. As the pipeline does not explicitly mention the construction of ontology, the knowledge engineering process does not adhere to the methodologies for creating an ontology and elicitation of knowledge from subject matter experts towards the creation of a knowledge model.

\textbf{Accessibility (Q08)} - KGTK is inspired by Scikit-learn and SpaCy, two popular toolkits for machine learning and natural language processing that have had a vast impact by making these technologies accessible to data scientists and software developers. Similarly, KGTK is a framework for manipulating, validating, and analyzing large-scale KGs that can be adopted for multi-purpose data-science operations over KGs, independently of the domain. Also, it is an open-source project and has well-documented documentation that helps users quickly get started working with KGTK. Hence, the barrier to adoption by users of the knowledge engineering process is low.

\textbf{Supports software engineering best practices (Q09)} - KGTK is an open-source project and it is closely tied to the level of adoption and contributions from the community. It supports features such as continuous integration and deployment, version control, and testing. Hence, the knowledge engineering process does conform to software industry norms (e.g. the use of agile methodologies, continuous integration and deployment, version control, automated testing, automated vulnerability scans, etc.)

\textbf{Robustness to noise (Q10)} - KGTK supports validate and clean operation which ensures the information that is added to the database at the time of data integration is valid and does not contain any potential errors. For example, the validate operation ensures that the data meets the KGTK file format specification, detecting errors such as nodes with empty values, values of unexpected length (either too long or too short), potential errors in strings (quotation errors, incorrect use of language tags, etc.), incorrect values in dates, etc. Similarly, the clean operation fixes a substantial number of errors detected by validation, by correcting some common mistakes in data encoding (such as not escaping ‘pipe’ characters), replacing invalid dates, normalizing values for quantities, languages, and coordinates using the KGTK convention for literals. Finally, it removes rows that still do not meet the KGTK specification (e.g., rows with empty values for required columns or rows with an invalid number of columns). Hence, the knowledge production process is robust in the face of noise and/or adversarial manipulation of source data and/or knowledge.

\textbf{Robustness to missing schema (Q11)} - KGTK handles missing schema using graph curations operations such as validate and clean. The validate operation ensures that the data meets the KGTK file format specification, detecting errors such as nodes with empty values. Similarly, the clean operation fixes a substantial number of errors detected by validating the operation and removes rows that still do not meet the KGTK specification (e.g., rows with empty values for required columns or rows with an invalid number of columns). Hence, the knowledge produced by the knowledge engineering process can be processed and/or accessed in the face of incomplete schemas and/or knowledge organization systems.

\textbf{Openness (Q12)} - KGTK is an open-source project i.e., all the modules are developed without any external dependencies and can be utilized by any developer to their fullest extent. It is being continuously updated and new features are getting added to it. It has well-documented documentation that enables users to quickly get started with the toolkit. Hence, the components of the knowledge engineering process are implemented using open-source software, with open standards, and the knowledge produced by the knowledge engineering process is openly accessible.

\textbf{Comprehensiveness (Q13)} - KGTK currently supports 13 operations grouped into 4 modules - importing modules, graph manipulation modules, graph analytics modules, and exporting modules. More specifically, it supports data ingestion, data integration, data transformation, inference, and data exporting. All in all, all components of an end-to-end knowledge engineering process (e.g. data ingest/export, data transformation, inference, knowledge publishing, etc.) are supported.

\textbf{Customizability (Q14)} - KGTK has illustrated usage by working with large open generic KGs such as Wikidata, DBpedia, and ConceptNet and domain-specific use cases such as the CORD-19 dataset. The heterogeneity of these cases shows how KGTK can be adopted for multi-purpose data-science operations over KGs, independently of the domain. In each of these use cases, the methods employed in the pipeline are modified to suit specific needs. KGTK has a pipelining architecture based on Unix pipes that allows chaining most operations in the required fashion using the input/output commands and KGTK file format. Hence, the components of the knowledge engineering process can be modified to support specific use cases.

\textbf{Modularity (Q15)} - The KGTK pipeline allows users to compose modules selectively depending upon the use case requirements. Additionally, it uses a simple representation file format that all modules in the toolkit operate on to enable tool integration without additional data transformations. It has the ability to integrate mature existing tools without a need for a new implementation, contains a comprehensive set of features - import/export, transformation, analytics, etc, and uses a pipeline mechanism that allows composing modules in arbitrary ways to process large KGs. Hence, the components of the knowledge engineering process can be selectively composed to suit a specific use case.

\textbf{Domain-independence (Q16)} - KGTK is a data science-centric approach designed to represent, create, transform, enhance, and analyze KGs. The authors have illustrated the importance of integrating and manipulating large open generic KGs such as Wikidata, DBpedia, and ConceptNet. Additionally, they used KGTK with the CORD-19 dataset provided by the Allen Institute for AI. They enhanced the dataset with KGs such as DBpedia, and Wikidata to incorporate gene, chemical, disease, and taxonomic information, and computing network analytics on the resulting graphs. Hence, the knowledge engineering process is generally applicable across a wide range of domains and areas of expertise.

\textbf{Extensibility (Q17)} - KGTK utilizes an import module to transform different input data formats into KGTK file format to ensure interoperability. One of the KGTK operations under the graph curation and transformation module is join which allows to integrate and extend the existing datasets with additional information. For example, the authors used KGTK with the CORD-19 dataset provided by the Allen Institute for AI. They enhanced the dataset with KGs such as DBpedia, and Wikidata to incorporate gene, chemical, disease, and taxonomic information, and computing network analytics on the resulting graphs. Hence, knowledge extraction from data or natural language performed in the knowledge engineering process can easily accommodate new sources and modalities of data or natural language.

\textbf{Curatability (Q18)} - Once the data is in KGTK format, curating operations can be performed such as cleaning and validation. The validate operation ensures that the data meets the KGTK file format specification, detecting errors such as nodes with empty values, values of unexpected length (either too long or too short), potential errors in strings (quotation errors, incorrect use of language tags, etc.), incorrect values in dates, etc. Similarly, the clean operation fixes a substantial number of errors detected by validate operation, by correcting some common mistakes in data encoding (such as not escaping ‘pipe’ characters), replacing invalid dates, normalizing values for quantities, languages, and coordinates using the KGTK convention for literals. Finally, it removes rows that still do not meet the KGTK specification (e.g., rows with empty values for required columns or rows with an invalid number of columns). As manual curation is not explicitly mentioned, the knowledge engineering process does not support the human curation of automatically extracted and/or inferred knowledge.

\textbf{Ethicality (Q19)} - As KGTK is an open-source project, it ensures legal compliance, collaborative and transparent development, community oversight, and quality control. These measures provide fairness, transparency, privacy, and accountability. Additionally, all the methods in all the modules are developed by the KGTK authors hence there is no dependency that might have affected transparency otherwise. Hence, the knowledge engineering process supports compliance with and enforcement of policies and/or guidelines for ethical use.

\textbf{Sustainability (Q20)} - KGTK is a data science-centric toolkit designed to represent, create, transform, enhance, and analyze large KGs. KGTK currently supports 13 operations grouped into 4 modules - importing modules, graph manipulation modules, graph analytics modules, and exporting modules. It does not involve manual development of ontology during the process nor does it require curating large sets of data for large language models (except embeddings which are optional) unlike the Tamašauskaitė and Groth knowledge graph development process where ML models were required in almost every step of the process. Hence, the cost of executing the knowledge engineering process is economically sustainable for the given use case.

\textbf{Affordability (Q21)} - KGTK is an open-source project and depending upon the use case, the start-up cost may vary, and it may incur costs for infrastructural setup, training, or support. Hence the cost of access to the knowledge engineering process is not economically affordable for a given user community.

\textbf{Editability (Q22)} - KGTK helps manipulate, curate, and analyze large real-world KGs. Throughout the pipeline it follows ways of manipulating and curating the information using operations such as validate and clean, it does not explicitly mention editing information manually by humans. Hence, the knowledge produced by the knowledge engineering process cannot be feasibly edited by humans.

\textbf{Explainability (Q23)} - KGTK follows a pipeline mechanism that allows composing modules in arbitrary ways to process large public KGs such as Wikidata, DBpedia, or ConceptNet which makes the entire process automated. Additionally, additional features can be passed alongside node1, label, and node2 feature columns in the KGTK file format to include provenance information about the creator of the statement and the original source. Additionally, the knowledge-creation process stores the intermediate outputs on disk, making it reproducible and easy to inspect. Hence, the knowledge produced by the knowledge engineering process provides accountability with respect to provenance and the details of how it was produced (e.g. through human authoring, automated extraction, and/or inference, etc.)










\end{document}